# A MIXTURE OF EXPERTS MODEL FOR RANK DATA WITH APPLICATIONS IN ELECTION STUDIES

By Isobel Claire Gormley[1] and Thomas Brendan Murphy[2]

*University College Dublin*

A voting bloc is defined to be a group of voters who have similar voting preferences. The cleavage of the Irish electorate into voting blocs is of interest. Irish elections employ a "single transferable vote" electoral system; under this system voters rank some or all of the electoral candidates in order of preference. These rank votes provide a rich source of preference information from which inferences about the composition of the electorate may be drawn. Additionally, the influence of social factors or covariates on the electorate composition is of interest.

A mixture of experts model is a mixture model in which the model parameters are functions of covariates. A mixture of experts model for rank data is developed to provide a model-based method to cluster Irish voters into voting blocs, to examine the influence of social factors on this clustering and to examine the characteristic preferences of the voting blocs. The Benter model for rank data is employed as the family of component densities within the mixture of experts model; generalized linear model theory is employed to model the influence of covariates on the mixing proportions. Model fitting is achieved via a hybrid of the EM and MM algorithms. An example of the methodology is illustrated by examining an Irish presidential election. The existence of voting blocs in the electorate is established and it is determined that age and government satisfaction levels are important factors in influencing voting in this election.

**1. Introduction.** The President of Ireland is elected every seven years by the Irish electorate through a preferential voting system known as the single

———————
Received October 2007; revised February 2008.

[1]Supported by a Government of Ireland Research Scholarship in Science, Engineering and Technology provided by the Irish Research Council for Science, Engineering and Technology, funded by the National Development Plan.

[2]Supported in part by a Science Foundation of Ireland Research Frontiers Programme Grant (06/RFP/M040).

transferable vote (STV). Under this system voters rank some or all of the presidential candidates in order of preference. An intricate vote counting process involving the elimination of candidates and the transfer of votes results in the election of one candidate as President.

A voting bloc is defined to be a group of voters who have similar voting preferences. The cleavage of any electorate into voting blocs is of interest to political scientists, politicians and voters. The cleavage of the Irish electorate is of particular interest, given the detailed, multi-preference votes expressed under the STV voting system. All the information expressed in the ranked preferences of the votes must be exploited in order to determine the true composition of the electorate. Further, the influence of social factors on the voting bloc membership of a voter is also of interest.

This work aims to establish the presence of voting blocs within the 1997 Irish presidential electorate, and to determine the characteristic voting preferences of these blocs. Additionally, the influence of social factors on the voting bloc memberships of voters is explored. The ranked nature of the voting data is modeled using the Benter model for rank data [Benter (1994)] and a mixture model of these distributions provides a model-based approach to clustering voters into voting blocs [Gormley and Murphy (2008a)]. Voting bloc membership probabilities are treated as multinomial logistic functions of the social factors (or covariates) associated with a voter. Such a mixture model in which the membership probabilities are functions of covariates is a mixture of experts model [Jacobs et al. (1991)]. Thus, a mixture of experts model for rank data is developed, thereby extending the mixture model for rank data developed in [Gormley and Murphy (2006, 2008a)] by including covariate information to aid the characterization of the mixture components.

Section 2 details the setting of the 1997 Irish presidential election and provides an example of the mechanics of the STV vote counting process. A mixture of experts model for rank data is formulated in Section 3 and unique model fitting aspects are discussed in Section 4. Model fitting is achieved via a hybrid of the popular EM algorithm [Dempster, Laird and Rubin (1977)] with the MM algorithm [Hunter and Lange (2004), Lange, Hunter and Yang (2000)] to produce an algorithm which we call the Expectation Minorization Maximization (EMM) algorithm. The mixture of experts model for rank data is fitted to the 1997 presidential electorate and the resulting model parameter estimates are discussed in Section 5. We compare the results in this analysis to other analyses of the data in Section 6. The article concludes with a discussion of the developed methodology.

**2. Irish presidential elections.** Irish presidential elections employ the Single Transferable Vote (STV) system. Under this system voters rank some or all of the electoral candidates in order of preference. The votes are totalled through a series of counts, where candidates are eliminated and their votes are transferred between candidates. An in-depth description of the electoral system, including the method of counting votes, is given in Sinnott (1999) and good introductions to the Irish political system are given in Coakley and Gallagher (1999) and Sinnott (1995). Further, an illustrative example of the manner in which votes are counted and transferred follows in Section 2.2. We start with a description of data from the 1997 presidential election in Section 2.1.

2.1. *The 1997 presidential election.* The current President of Ireland, Mary McAleese, is in her second term of office. Originally elected in 1997, she was automatically re-elected in 2004 as the only validly nominated candidate.

In the 1997 presidential election there were five candidates: Mary Banotti, Mary McAleese, Derek Nally, Adi Roche and Rosemary Scallon. As detailed in Table 2, some candidates were endorsed by political parties and others were independent candidates. Mary McAleese had a high public profile and received the backing of Fianna Fáil, who were the main political party in the coalition government at the time. Mary Banotti was another high profile candidate who was endorsed by the main government opposition party, Fine Gael. Adi Roche was supported by the Labour party, who were also a government opposition party. The remaining two candidates ran on independent tickets. In terms of campaign themes, Mary Banotti, Derek Nally and Adi Roche were considered to be liberal candidates, whereas Mary McAleese and Rosemary Scallon were deemed more conservative candidates. A detailed description of the entire presidential election campaign, including the nomination and selection of candidates, is given in Marsh (1999).

An opinion poll conducted by Irish Marketing Surveys one month prior to the election is analyzed in this article. Interviews were conducted on 1100 respondents, drawn from 100 sampling areas. Interviews took place at randomly located homes, with respondents selected according to a socioeconomic quota. A range of sociological questions was asked of each respondent, as was their voting preference, if any, for each of the candidates. These preferences were utilized as a statement of the intended voting preferences of each respondent. Of the respondents interviewed, 17 indicated that they did not intend to vote—these respondents were excluded from the analysis. Table 1 details the set of sociological covariates recorded in the poll.

2.2. *The vote counting process.* A brief overview of the vote counting process is given here. For illustrative purposes, the transfer of votes in the 1997 Irish presidential election is shown in Table 2.

Under the STV electoral system, a "quota" of votes is calculated which is dependent on the number of seats available and the number of valid votes cast. Specifically, the quota is computed as

$$\frac{\text{total valid votes cast}}{\text{number of seats to be filled} + 1} + 1.$$

Thus, for the 1997 presidential election, where 1,269,836 valid votes were cast and a single presidential seat was to be filled, the quota was calculated to be 634,919. Once any candidate at any counting stage obtained or exceeded 634,919 votes, this candidate was deemed elected as President of Ireland. As detailed in Table 2, in the first stage of the counting process the number of first preference votes obtained by each candidate is totaled. No candidate received enough first preference votes to exceed the quota. Mary

TABLE 1

*The set of covariates recorded in the presidential election opinion poll and the associated levels (in the case of categorical variables). An explanation of the socioeconomic group codes are provided in Appendix B*

| Age | Area | Gender | Government satisfaction | Marital status | Socioeconomic group |
|---|---|---|---|---|---|
| — | City | Housewife | Satisfied | Married | AB |
| | Town | Nonhousewife | Dissatisfied | Single | C1 |
| | Rural | Male | No opinion | Widowed | C2 |
| | | | | | DE |
| | | | | | F50+ |
| | | | | | F50− |

TABLE 2

*The transfer of votes in the 1997 presidential election. The quota required to be elected President of Ireland was 634,919. Mary McAleese (denoted in bold font) was elected*

| Candidate | Endorsing party | Count 1 | Count 2 |
|---|---|---|---|
| Mary Banotti | Fine Gael | 372,002 | *+125,514* |
| | | | *497,516* |
| **Mary McAleese** | **Fianna Fáil** | **574,424** | **+131,835** |
| | | | **706,259** |
| Derek Nally | Independent | 59,529 | *−59,529* |
| | | | *Eliminated* |
| Adi Roche | Labour | 88,423 | *−88,423* |
| | | | *Eliminated* |
| Rosemary Scallon | Independent | 175,458 | *−175,458* |
| | | | *Eliminated* |
| Nontransferable votes | | | *+66,061* |
| | | | 66,061 |
| Total valid votes | | 1,269,836 | 1,269,836 |

McAleese received the largest number of first preference votes with 45% of the vote share. Candidates Nally, Roche and Scallon were eliminated from the election race after the first count, as the sum of their votes was less than the votes of the next lowest candidate (Mary Banotti).

At the second stage of counting, Nally, Roche and Scallon's 323,410 first preference votes were transferred to the candidates given the next valid preference on those ballot papers. Of votes to be transferred, 66,061 were nontransferable because only a single preference was expressed on these ballots or lower preferences on the ballots were for eliminated candidates. Mary McAleese received 131,835 of the transferred votes and was therefore elected at the second counting stage, as she exceeded the quota with 706,259 votes.

**3. A mixture of experts model for rank data.** The mixture of experts (MoE) model [Jacobs et al. (1991), Jordan and Jacobs (1994)] combines the ideas of mixture models [McLachlan and Peel (2000)] and generalized linear models [McCullagh and Nelder (1983), Dobson (2002)]. A mixture model is used to model the heterogeneous nature of a population; generalized linear model theory provides the statistical structure within the mixture.

MoE models account for the relationship between a set of response and covariate variables where it is assumed that the conditional distribution of the response given the covariates is a finite mixture distribution. The conditional probability of voter $i$'s ballot $\underline{x}_i$, given their associated covariates $\underline{w}_i$, is

$$\mathbf{P}(\underline{x}_i|\underline{w}_i) = \sum_{k=1}^{K} \pi_{ik}\mathbf{P}(\underline{x}_i|\theta_k),$$

where $K$ denotes the number of components (or *expert networks*) in the mixture, the *gating network coefficient* $\pi_{ik} = \pi_k(\underline{w}_i)$ is the probability of voter $i$ being a complete member of expert network $k$ and $\theta_k$ represents the parameters of the probability model of the $k$th expert network. In the current context, an expert network corresponds to a voting bloc in the electorate. A more general mixture of experts model would allow the expert network parameters to depend on the covariates $\underline{w}_i$, however, such a model would be difficult to interpret in terms of voting blocs.

The gating network coefficients are weighting probabilities constrained such that they are nonnegative and sum to one for each voter. The probability of voter $i$'s ballot according to the expert networks in the mixture model are blended by the gating network coefficients to produce an overall probability. Thus, the probability of voter $i$'s ballot is a convex combination of the output probabilities from the expert networks. Figure 1 provides a graphical illustration of the structure of a single layer MoE model—a hierarchical MoE model consists of multiple layers of expert networks and gating networks.

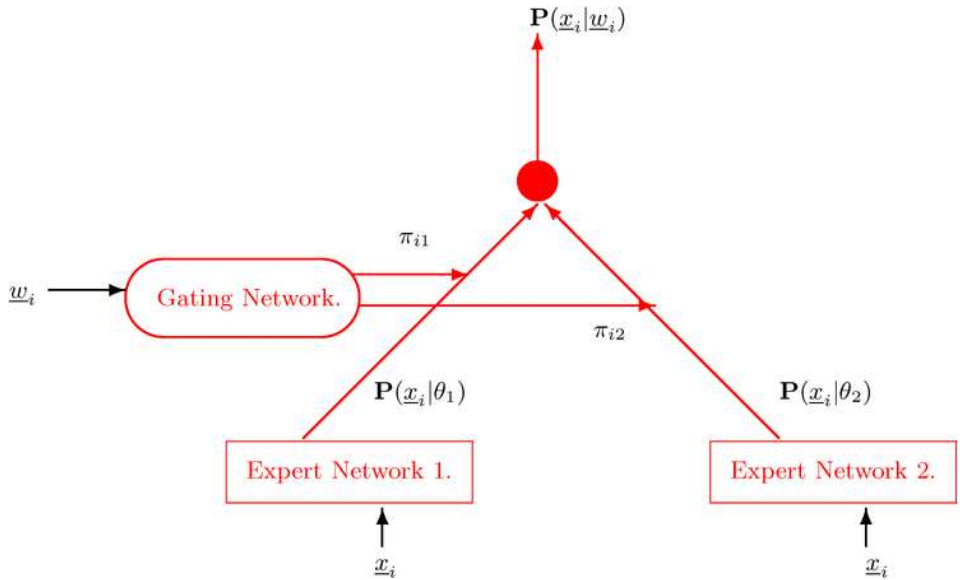

Fig. 1. *Graphical illustration of the structure of a single layer mixture of experts model with two expert networks. The probabilities of voter $i$'s ballot according to the expert networks $\mathbf{P}(\underline{x}_i|\theta_1)$ and $\mathbf{P}(\underline{x}_i|\theta_2)$ are blended by the gating network coefficients $\pi_{i1}$ and $\pi_{i2}$ to produce an overall probability of voter $i$'s ballot. The gating network coefficients are assumed to be a function of voter $i$'s covariates $\underline{w}_i$.*

Traditional MoE models, such as those fitted in Jordan and Jacobs (1994), Jacobs et al. (1991) and Peng, Jacobs and Tanner (1996), employ probability densities for the expert networks which are members of the exponential family, that is, the traditional MoE model has the form of a mixture of generalized linear models. In the context of STV voting data, however, the expert network probability densities must appropriately model the ranked nature of the data. Thus, the Benter model for rank data [Benter (1994)] is employed; full details are provided in Section 3.1.

As illustrated in Figure 1, the gating network coefficients are assumed to be functions of the voter covariates. The intuition here is that the covariates of a voter determine their voting bloc membership and, in turn, their characteristic voting preferences. Specifically, the gating network coefficients are assumed to be multinomial logistic functions of the voter covariates; details are provided in Section 3.2.

The tree-like structure of MoE models naturally induces comparisons to other tree-based classification methods, such as Classification and Regression

Trees (CART) [Breiman et al. (2006)] or Multivariate Adaptive Regression Splines (MARS) [Friedman (1991)]. Both CART and MARS are nonparametric techniques which provide a hard partition of the data space—using these tools each voter would be classified as belonging to one and only one voting bloc. In contrast, the statistical models underlying the expert network probability densities mean MoE models are parametric in nature. Additionally, the MoE model provides a probabilistic "soft" partition of the space in that data points may belong to multiple expert networks, that is, under the MoE model each voter has an associated probability of belonging to each voting bloc. Further comparison of these methods is provided in Peng, Jacobs and Tanner (1996) and Bishop (2006).

3.1. *Benter's model for rank data.* In previous versions of the MoE model [Jacobs et al. (1991), Jordan and Jacobs (1994), Peng, Jacobs and Tanner (1996)] it is assumed that each component of the mixture model (i.e., each expert network) produces its output as a generalized linear function of input predictor variables. Within the context of STV voting data, each expert network must appropriately model the ranked nature of the data. Thus, it is assumed that each expert network is a Benter model distribution for rank data [Benter (1994)]. Each expert network is characterized by a differently parameterized Benter model where the parametrization is constant with respect to the voter covariates. It would be possible to allow covariates to contribute to the expert networks [see Jordan and Jacobs (1994) and Peng, Jacobs and Tanner (1996)], but this is not examined here due to the fact that interpreting the expert networks in terms of voting blocs would be difficult.

The Benter model for rank data has two parameters—a *support parameter* and a *dampening parameter*:

(i) *Support parameter.* Within expert network $k$, the support parameter vector is denoted $\underline{p}_k = (p_{k1}, \ldots, p_{kN})$, where $0 \leq p_{kj} \leq 1$, $\sum_{j=1}^{N} p_{kj} = 1$ and $N$ denotes the number of candidates available for selection. The support parameter $p_{kj}$ may be interpreted as the probability of candidate $j$ being given a first preference by a complete member of voting bloc $k$.

(ii) *Dampening parameter.* The global dampening parameter vector is denoted by $\underline{\alpha} = (\alpha_1, \ldots, \alpha_N)$, where $\alpha_t \in [0, 1]$ for $t = 1, \ldots, N$. To avoid overparametrization of the model, the constraints $\alpha_1 = 1$ and $\alpha_N = 0$ are imposed. The dampening parameters model the way in which some preferences may be chosen less carefully than other preferences within a ballot.

Let $c(i, t)$ denote the candidate ranked in $t$th position by voter $i$ and $n_i$ be the total number of preferences expressed by voter $i$. Given the Benter model parameters, the probability of voter $i$'s ballot (conditional on voter $i$ being a complete member of voting bloc $k$) is

$$\mathbf{P}(\underline{x}_i|\theta_k) = \mathbf{P}(\underline{x}_i|\underline{p}_k, \underline{\alpha}) = p_{kc(i,1)}^{\alpha_1} \cdot \frac{p_{kc(i,2)}^{\alpha_2}}{\sum_{s=2}^{N} p_{kc(i,s)}^{\alpha_2}} \cdots \frac{p_{kc(i,n_i)}^{\alpha_{n_i}}}{\sum_{s=n_i}^{N} p_{kc(i,s)}^{\alpha_{n_i}}}$$

(3.1)

$$= \prod_{t=1}^{n_i} \frac{p_{kc(i,t)}^{\alpha_t}}{\sum_{s=t}^{N} p_{kc(i,s)}^{\alpha_t}}.$$

Thus, the Benter model states that the probability of a rank ballot is the product of the probabilities of each chosen candidate being ranked first where, at each preference level, the probabilities are appropriately normalized to account for the fact that the cardinality of the choice set has been reduced. Moreover, at preference level $t$, the care with which a preference is made is modeled by "dampening" each probability by $\alpha_t$.

Under the Benter model, the log odds of selecting candidate $a$ over candidate $b$ at preference level $t$ is $\alpha_t \log(p_{ka}/p_{kb})$. Thus, the $t$th level dampening parameter $\alpha_t$ can be interpreted as how the log odds of selecting candidate $a$ over candidate $b$ is affected by the selection being made at preference level $t$.

The Benter model has been successfully employed to model rank data [see Gormley and Murphy (2006, 2008a)], but alternative rank data models are also available; the Plackett–Luce model for rank data [Plackett (1975)] is a special case of the Benter model in which the dampening parameter vector is constrained such that $\underline{\alpha} = \underline{\mathbf{1}}$. Under the Plackett–Luce model, it is assumed that a voter makes their choice at each preference level with equal certainty. Mixtures of Plackett–Luce models have been fitted to rank data in Gormley and Murphy (2006) and the fitting algorithms for these models are more efficient due to the fixed $\underline{\alpha}$ value. The Benter and Plackett–Luce models are both *multistage ranking models* [Marden (1995)]; in Fligner and Verducci (1986) this large class of models are defined as those which decompose the ranking process into a series of independent stages. Such models have an "item-effect" approach [Fligner and Verducci (1986)] in that the probability of the preference of one item over another is the element of interest. Another set of rank data models are "distance" based such as those based on Mallow's model [Mallows (1957)]; in such models the probability of observing a ranking $\underline{x}$ decreases as the distance between $\underline{x}$ and the modal ranking $\underline{y}$ increases. Other distance based approaches are detailed in Gordon (1979) and Fligner and Verducci (1986). Cluster analysis via mixtures of distance based models is described in Murphy and Martin (2003) and Busse et al. (2007). Given the type of choice process undertaken by a voter when generating an STV ballot paper, the Benter model for rank data was deemed the most applicable in this context.

3.2. *Gating network coefficients and generalized linear models.* The gating network coefficients in the MoE model can be viewed as the success probabilities from a generalized linear model. In particular, the success probability of belonging to each of $K$ expert networks is a multinomial logistic function of the covariates (see Figure 1). Voter $i$'s gating network coefficients $\underline{\pi}_i = (\pi_{i1}, \pi_{i2}, \ldots, \pi_{iK})$ are modeled by a logistic function of their $L$ covariates $\underline{w}_i = (w_{i1}, w_{i2}, \ldots, w_{iL})$, that is,

$$(3.2) \qquad \log\left(\frac{\pi_{ik}}{\pi_{i1}}\right) = \beta_{k0} + \beta_{k1}w_{i1} + \beta_{k2}w_{i2} + \cdots + \beta_{kL}w_{iL},$$

where expert network 1 is used as the baseline expert network and $\beta_{k0}$ is an intercept term. Similar methodology was employed in Jordan and Jacobs (1994) and Peng, Jacobs and Tanner (1996) when modeling the gating network coefficients.

## 4. Fitting the MoE model via the EMM algorithm.
To determine the composition and voting characteristics of the Irish electorate, estimates of the Benter model parameters and of the gating network coefficients are required. Model fitting of the MoE model is achieved in Jacobs et al. (1991) and Jordan and Jacobs (1994) via the Expectation Maximization (EM) algorithm. Estimation of the MoE model within the Bayesian framework is detailed in Peng, Jacobs and Tanner (1996), in which Markov chain Monte Carlo methods [Tanner (1996)] are used. An alternative approach to estimation of the MoE parameters within the Bayesian framework through the use of variational methods is detailed in Bishop and Svensén (2003).

In this article parameter estimation is achieved through a hybrid algorithm known as the "EMM" algorithm. As the name implies, the EMM algorithm combines the well-known EM algorithm [Dempster, Laird and Rubin (1977)] with ideas from the MM algorithm [Lange, Hunter and Yang (2000)].

4.1. *The EM algorithm for the MoE model.* The EM algorithm is most commonly known as a technique to produce maximum likelihood estimates (MLEs) of model parameters in settings where the data under study is incomplete or when optimization of the likelihood would be simplified if an additional set of variables were known. The iterative EM algorithm consists of an expectation (E) step followed by a maximization (M) step. Generally, during the E step the expected value of the log likelihood of the complete data (i.e., the observed and unobserved data) is computed. In the M step the expected log likelihood is maximized with respect to the model parameters. In practice, the imputation of latent variables often makes maximization of the expected log likelihood feasible. The parameter estimates produced in the M step are then used in a new E step and the cycle continues until convergence. The parameter estimates produced on convergence are estimates which achieve at least a local maximum of the likelihood function of the data.

It is difficult to directly obtain MLEs from the likelihood of the MoE model for $M$ rank observations:

$$\mathcal{L}(\boldsymbol{\beta}, \mathbf{p}, \underline{\alpha} | \mathbf{x}, \mathbf{w}) = p(\mathbf{x} | \mathbf{w}, \boldsymbol{\beta}, \mathbf{p}, \underline{\alpha}) = \prod_{i=1}^{M} \sum_{k=1}^{K} \pi_{ik}(\underline{w}_i) \mathbf{P}(\underline{x}_i | \underline{p}_k, \underline{\alpha}).$$

To alleviate this problem, the data is augmented by imputing latent variables. For each voter $i = 1, \ldots, M$, the latent variable $\underline{z}_i = (z_{i1}, \ldots, z_{iK})$ is imputed where $z_{ik}$ takes the value 1 if voter $i$ is a complete member of expert network $k$ and the value 0 otherwise. This provides the complete data likelihood

$$\mathcal{L}_C(\boldsymbol{\beta}, \mathbf{p}, \underline{\alpha} | \mathbf{x}, \mathbf{z}, \mathbf{w}) = p(\mathbf{x}, \mathbf{z} | \mathbf{w}, \boldsymbol{\beta}, \mathbf{p}, \underline{\alpha}) = \prod_{i=1}^{M} \prod_{k=1}^{K} \{\pi_{ik}(\underline{w}_i) \mathbf{P}(\underline{x}_i | \underline{p}_k, \underline{\alpha})\}^{z_{ik}},$$

the expectation of (the log of) which is obtained in the E step of the EM algorithm. Details are provided in Appendix C, but, in brief, the E step consists of replacing the missing data $\mathbf{z}$ with their expected values $\hat{\mathbf{z}}$. In the M step the complete data log likelihood, computed with the estimates $\hat{\mathbf{z}}$, is maximized to provide estimates of the Benter parameters $\hat{\mathbf{p}}$ and $\underline{\hat{\alpha}}$ and the gating network parameters $\hat{\beta}$.

The EM algorithm for fitting the MoE model for rank data is straightforward in principle, but the M step is difficult in practice. This is largely due to the complex form of the Benter model density (3.1) and the large parameter set. A modified version of the EM algorithm, the Expectation and Conditional Maximization (ECM) algorithm [Meng and Rubin (1993)], is therefore employed. In the ECM algorithm, the M step consists of a series of conditional maximization steps. Again, in the context of the MoE model for rank data, these maximizations are not straight forward and, thus, the conditional M step is implemented using the MM algorithm.

4.2. *The MM algorithm.* The MM algorithm is a summary term for optimization algorithms which operate by transferring optimization from the objective function of interest to a more tractable surrogate function. Good summaries of the methodology are provided in Lange, Hunter and Yang (2000), Hunter and Lange (2004) and Hunter (2004). The initials MM depend on the type of optimization required. In a maximization problem MM stands for minorize and maximize; in a minimization problem, majorize and minimize. A minorizing (or majorizing) surrogate function is constructed by exploiting mathematical properties of the objective function or of terms within it. The MM philosophy is that iteratively optimizing a suitable surrogate function drives the objective function uphill or downhill as required. Iteratively maximizing a minorizing surrogate function produces a sequence of parameter estimates which converges to at least a local maximum of the objective function. A graphical illustration of the mechanics of the MM algorithm is given in Figure 2. Details of the stability of MM algorithms and their relation to the EM algorithm (the EM algorithm is in fact an MM algorithm) are detailed in Lange, Hunter and Yang (2000) and Hunter and Lange (2004).

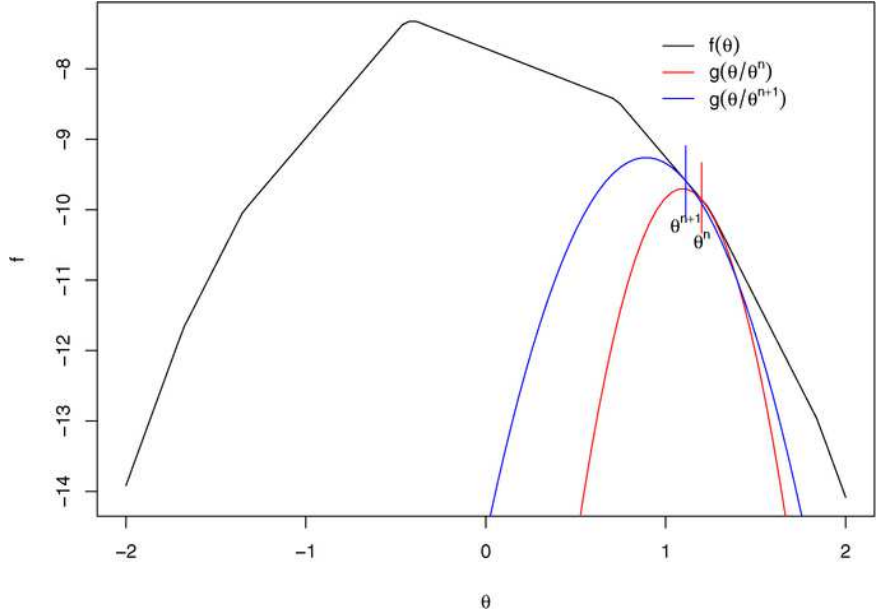

FIG. 2. *A graphical illustration of one iterative step in a maximization MM algorithm. A minorizing surrogate function $g(\theta|\theta^n)$ (in red) is first fitted to the objective function $f(\theta)$ (in black) at the parameter value $\theta^n$. Maximizing this minorizing surrogate function provides a new parameter estimate $\theta^{n+1}$. A new minorizing surrogate function is fitted (in blue) to the objective function at $\theta^{n+1}$. The process continues, driving the objective function uphill, until the parameter estimates converge, indicating that at least a local maximum of the objective function has been reached.*

In the context of fitting MoE models for rank data, the optimization problems in the conditional M step of the EM algorithm are overcome by embedding several iterations of the MM algorithm in place of the conditional M step. Details of the construction of the necessary surrogate functions are provided in Appendix C.

## 5. The MoE model for rank data and the Irish electorate.

The MoE model for rank data was fitted to the set of voters polled in the Irish Marketing Surveys opinion poll detailed in Section 2.1. For reasons of numerical stability and ease of interpretation, covariates were initially standardized such that $0 \leq w_{il} \leq 1$, where $w_{il}$ denotes the value of the $l$th covariate for voter $i$. A single layer MoE model rather than a hierarchical model was assumed to be sufficient in this context due to the small number of candidates in the presidential race.

Within a single layer MoE model, the number $K$ of expert networks (or voting blocs) present in the electorate needs to be estimated. In Jordan and Jacobs (1994) $K$ is chosen to be the value which minimizes a test set error rate; the variational Bayes approach taken in Bishop and Svensén (2003) provides a framework in which both the number of expert networks in and the topology of the MoE model may be estimated. The Bayesian Information Criterion (BIC) [Schwartz (1978)] is utilized here to select the optimal number of experts. The BIC is an information criterion motivated by the aim of minimizing the Kullback–Leibler information of the true model from the fitted model. The usual justification for the BIC is that, for regular problems, it is an approximation of the Bayes factor for comparing models under certain prior assumptions [Kass and Raftery (1995)]. The BIC is defined as

$$(5.1) \quad \text{BIC} = 2(\text{maximized likelihood}) - (\text{number of parameters}) \log(M),$$

where $M$ is the total number of data points. The BIC trades off model fit [assessed by the first term in (5.1)] against model complexity [assessed by the second term in (5.1)]. The use of BIC for model selection is not completely accepted; Gelman and Rubin (1995) and Raftery (1995) provide two contrasting views. Although mixture models do not satisfy the conditions necessary for the Bayes factor approximation to hold, there is much in the literature to support its use in this context [see, e.g., Leroux (1992), Keribin (1998), Keribin (2000) and Fraley and Raftery (1998)]. Within the context of this application, BIC gives reasonable results in terms of the voting blocs found.

As with any iterative procedure, starting values may be influential on the output of the algorithm. Starting values for the Benter support parameters, dampening parameters and missing membership labels were obtained by initially running the EMM algorithm for 500 iterations for a straight forward mixture of Benter models (i.e., the gating network coefficients are not treated as functions of the voter covariates). Good starting values for the gating network parameters were then obtained by running 1000 of the logistic regression style M steps [see (C.5)] of the EMM algorithm. The full EMM algorithm to provide MLEs of the model parameters was then iterated until convergence as deemed by Aitken's acceleration criterion [Böhning et al. (1994)]. Subsequent to convergence, approximate standard errors of the MLEs were calculated as detailed in McLachlan and Krishnan (1997) and McLachlan and Peel (2000).

Table 3

*The five best fitting MoE models as deemed by the BIC. Larger BIC values indicate better fitting models. The number of expert networks $K$ and the associated covariates of the models are also reported*

| BIC | K | Covariates |
|---|---|---|
| $-8490.43$ | 4 | Age<br>Government satisfaction |
| $-8498.59$ | 3 | Age |
| $-8507.33$ | 3 | Age<br>Government satisfaction |
| $-8511.37$ | 3 | Government satisfaction |
| $-8512.62$ | 5 | Age<br>Government satisfaction |

The MoE model was fitted over the range $K = 1, 2, \ldots, 5$ expert networks using a backward elimination style method to choose the informative covariates. Interaction terms were avoided. A model with all six covariates was initially fitted, then models with only five of the covariates. From this set of models the "best" model as deemed by the BIC was selected and models with only four of the selected covariates were then fitted. This selection of the best subset of covariates and then backward elimination was continued until only one covariate was left in the model. The BIC values for all the fitted models were then compared. Table 3 details the five best fitting models as deemed by their BIC values. The covariates within each selected model are also detailed.

Each of the five best fitting models considered age and/or government satisfaction as important covariates. The optimal model with $K = 4$ expert networks where age and government satisfaction are the influential covariates is discussed below.

5.1. *Benter support parameter estimates.* Figure 3 is a mosaic plot illustrating the Benter support parameter estimates within each of the four voting blocs in the optimal model. The voting blocs are each represented by a column and their associated marginal membership probabilities are reported.

Voting bloc 1 appears to favor the conservative candidates of McAleese and Scallon. The opinion poll was conducted at an early stage of the electoral campaign and Scallon had not yet established herself as a main presidential contender. Thus, the 31% support for Scallon in this voting bloc is the largest support she obtains. Voting bloc 2 also reveals characteristics of the early stages of the presidential campaign. Adi Roche has large support in this voting bloc—at the start of the campaign Roche was a very popular candidate, but her support quickly dropped when she became embroiled in difficulties and her campaign went into decline. Voting bloc 3, the largest voting bloc in terms of marginal membership probabilities, has a large support parameter for Mary McAleese. McAleese, who was subsequently elected, was backed by the current governmental political party, Fianna Fáil, and thus, she had a high public profile. There is also some level of support for the other high profile candidate, Mary Banotti. Voting bloc 4 is a pro-Banotti voting bloc with more uniform levels of support for the other candidates. Of note are the low levels of support for Nally in all of the voting blocs—Nally joined the electoral campaign later than the other candidates on September 29th and so had little time to win support prior to this October 2nd poll.

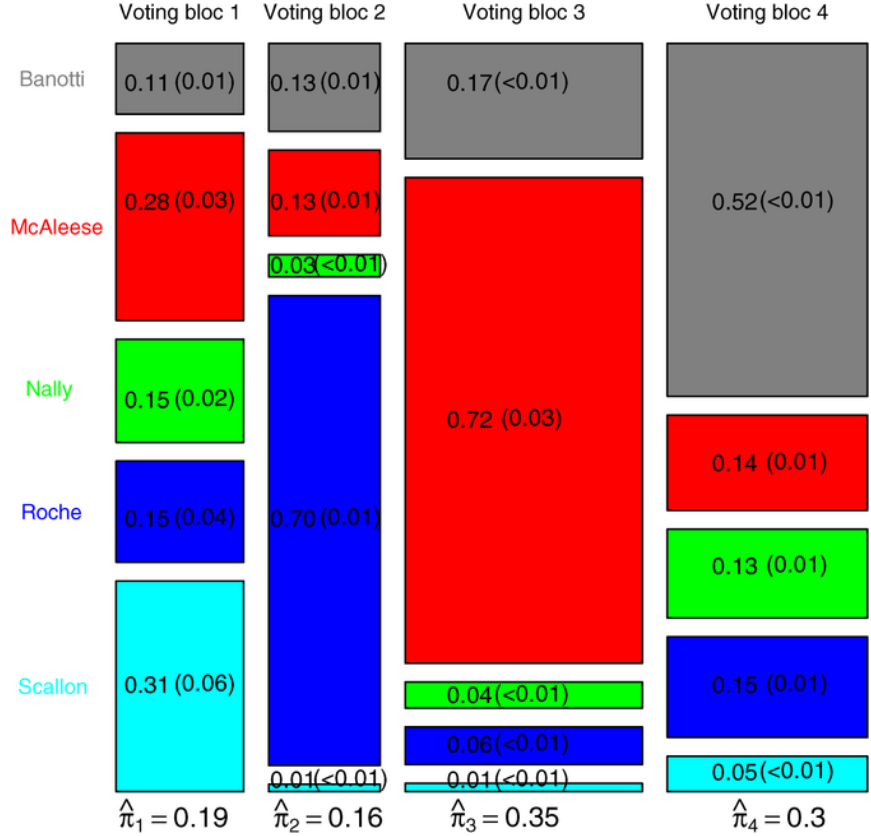

FIG. 3.  *A graphical representation of the maximum likelihood estimates of the Benter support parameters for the Irish Marketing Surveys opinion poll. Each column of the mosaic represents an expert network or voting bloc—the segments within the columns represent the magnitudes of the support parameters for the candidates within each voting bloc. The maximum likelihood estimates of the support parameters are detailed within each segment; standard errors are provided in parentheses. The width of each column represents the marginal probability $\pi_k$ of belonging to each voting bloc $k$.*

5.2. *Benter dampening parameter estimates.* Under the optimal model, the Benter dampening parameter estimates are

$$\hat{\underline{\alpha}} = (1.00, 0.99(0.10), 0.97(0.12), 0.99(0.15), 1.00)$$

(standard errors are given in parentheses). The estimates suggest that the certainty with which voters rank their preferences remains constant with respect to choice level. The proximity of the dampening parameter estimates to 1, along with their relatively large standard errors, suggest a Plackett–Luce model (Section 3.1) would be adequate for modeling this poll data.

The Benter dampening parameters appear to depend somewhat on the cardinality of the choice set; in this case, where the choice set is small, $\underline{\alpha} \approx \underline{1}$. In Gormley and Murphy (2008a) the Benter model is employed when modeling a larger choice set and $\underline{\alpha}$ is shown to differ from $\underline{1}$. Intuitively, the certainty associated with the ranking of objects from a small choice set would be greater than that associated with the ranking of objects from a large choice set.

5.3. *Gating network parameter estimates.* Under the MoE model for rank data, the gating network coefficients are functions of voter covariates. According to the BIC (see Table 3), the "age" and "government satisfaction" covariates influence the voting bloc membership probabilities of a voter. Table 4 details the associated gating network parameter estimates, their odds ratios and the relevant 95% confidence intervals for the odds ratios. The gating network parameters associated with the "conservative" voting bloc (i.e., voting bloc 1) are used as the reference parameters, that is, $\underline{\beta}_1 = (\beta_{10}, \ldots, \beta_{1L}) = (0, \ldots, 0)$.

Within the smallest voting bloc (i.e., voting bloc 2 or the pro-Roche bloc), for every one unit increase in age the odds for being best described by voting bloc 2 are 100 times less than the odds for being described by the conservative voting bloc 1. This would appear to be an intuitive characteristic of the Irish electorate—the more elderly generations in Ireland would, in general, be considered more conservatively minded. Note also the relatively small associated odds ratio confidence interval. The 95% confidence intervals for the government satisfaction covariate odds ratios both enclose 1, implying it is likely that the political views of voters in this bloc have little influence. Thus, younger voters appear to be best described by voting bloc 2 and are more in favor of the liberal Adi Roche.

In terms of the gating network parameters which refer to voting bloc 3 (the pro-McAleese bloc), the confidence interval for the age parameter odds

TABLE 4

*Gating network parameter estimates $\hat{\underline{\beta}}_k$, the associated odds ratios and the 95% odds ratio confidence intervals under the MoE model for rank data fitted to the Irish Marketing Surveys opinion poll data. The covariates selected as informative are age and government satisfaction.*

*"Do not know/no opinion" was used as the reference level within the categorical government satisfaction covariate*

| | | Intercept | Age | Satisfied | Not satisfied |
|---|---|---|---|---|---|
| Voting bloc 2 | Log odds ($\hat{\underline{\beta}}_2$) | 0.92 | −5.16 | 0.13 | 1.03 |
| | Odds ratio [exp($\hat{\underline{\beta}}_2$)] | 2.52 | 0.01 | 1.14 | 2.80 |
| | 95% CI (Odds ratio) | [0.78, 8.16] | [0.00,0.05] | [0.42, 3.11] | [0.77, 10.15] |
| Voting bloc 3 | Log odds ($\hat{\underline{\beta}}_3$) | −0.46 | −0.05 | 1.14 | 1.33 |
| | Odds ratio [exp($\hat{\underline{\beta}}_3$)] | 0.63 | 0.95 | 3.12 | 3.81 |
| | 95% CI (Odds ratio) | [0.16, 2.49] | [0.32,2.81] | [0.94, 10.31] | [0.90, 16.13] |
| Voting bloc 4 | Log odds ($\hat{\underline{\beta}}_4$) | 0.54 | 0.44 | −1.05 | 1.25 |
| | Odds ratio [exp($\hat{\underline{\beta}}_4$)] | 1.71 | 1.56 | 0.35 | 3.50 |
| | 95% CI (Odds ratio) | [0.52, 5.58] | [0.35, 6.91] | [0.12, 0.98] | [1.07, 11.43] |

ratio includes 1, suggesting age is not a driving covariate. The government satisfaction covariate appears to be more influential: the odds of a voter being best described by voting bloc 3 are around 3 times greater than the odds for voting bloc 1, given that the voter has some political opinion. Thus, voters with an interest in politics appear to favor Mary McAleese.

The gating parameters for voting bloc 4 indicate that voters with a dislike for the current government favor Mary Banotti. The confidence interval for the age covariate again includes 1, suggesting it has little effect. The odds of a voter who indicated a dislike for the 1997 government (a coalition government of Fianna Fáil and the Progressive Democrats) being best described by voting bloc 4 were 3.50 times greater than being described by voting bloc 1. In contrast, the odds of a voter in favor of the current government being best described by voting bloc 4 are 0.35 times greater than the odds for voting bloc 1. These results make intuitive sense within the context of the 1997 presidential election. Mary Banotti was endorsed by Fine Gael, the main opposition party to Fianna Fáil. Thus, voters best described by voting bloc 4 appear to be Fine Gael supporters. Those voters in favor of the 1997 coalition government were more likely to be described by voting bloc 1, which had large levels of support for Fianna Fáil backed McAleese.

## 6. Comparison.

6.1. *Results.* The analysis completed here is an extension of previous work exploring voting blocs in Irish elections [Gormley and Murphy (2008a)].

The mixture of experts model provides an extension of the mixture model because it allows us to assess which social factors influence voting bloc membership. The mixture of experts analysis suggests that "age" and "government satisfaction" influence voting bloc membership.

The analysis presented in Gormley and Murphy (2008a), for the same opinion poll presented here, finds just two voting blocs in the electorate. A mosaic plot illustrating the support parameters in the two voting blocs is given in Figure 4. The first voting bloc is a "noise" component where each candidate has equal support; such a component can collect small voting blocs and voters with unusual preference patterns. In this case, voting bloc 1 and, to an extent, voting bloc 4 from the mixture of experts model (Figure 3) are being combined in the noise group. The second voting bloc in Figure 4 is very similar to voting bloc 3 in the mixture of experts analysis, but also contains some of the voters from voting blocs 2 and 4. It is not surprising that more voting blocs are found using the mixture of experts model, because voting bloc membership is estimated from covariates recording social factors and voting behavior, whereas in the mixture model we only have voting data. Hence, the mixture of experts model is able to exploit the more detailed structure within the electorate than the standard mixture model can.

A comparison of the voting blocs found in the analysis of an opinion poll one month prior to the election (as presented here) with the voting blocs found for the exit poll [as shown in Figure 2 in Gormley and Murphy (2008a)] shows very different results. Noticeably, voting bloc 2 in the mixture of experts analysis is not present in the exit poll analysis—this is because support for Roche collapsed in the intervening month. However, both polls show voting blocs that have strong support for McAleese and Banotti respectively.

A latent space model was used in Gormley (2006) and Gormley and Murphy (2007) to model rank data and, in particular, Irish election data. They showed that the 2002 Irish general election and the 1997 presidential elections can be modeled using a one or two-dimensional latent space. In particular, the opinion poll considered in this paper was analyzed in Gormley (2006) and it was found that the election could be modeled using a one or two-dimensional latent space. The low dimensionality of the latent space found in Gormley (2006) is mirrored in the small number of voting blocs found in the mixture model in Gormley and Murphy (2008a) and the mixture of experts model here.

An analysis of the opinion poll data from the presidential election presented in van der Brug, van der Eijk and Marsh (2000) showed that a large number of voters' first preference candidate coincided with their first preference national party, so the election had a partisan aspect to it. However, they suggest that the election was not strongly partisan. Our analysis gives similar results in suggesting that support for the government was an important aspect in the membership of voting blocs but that voter age was also influential.

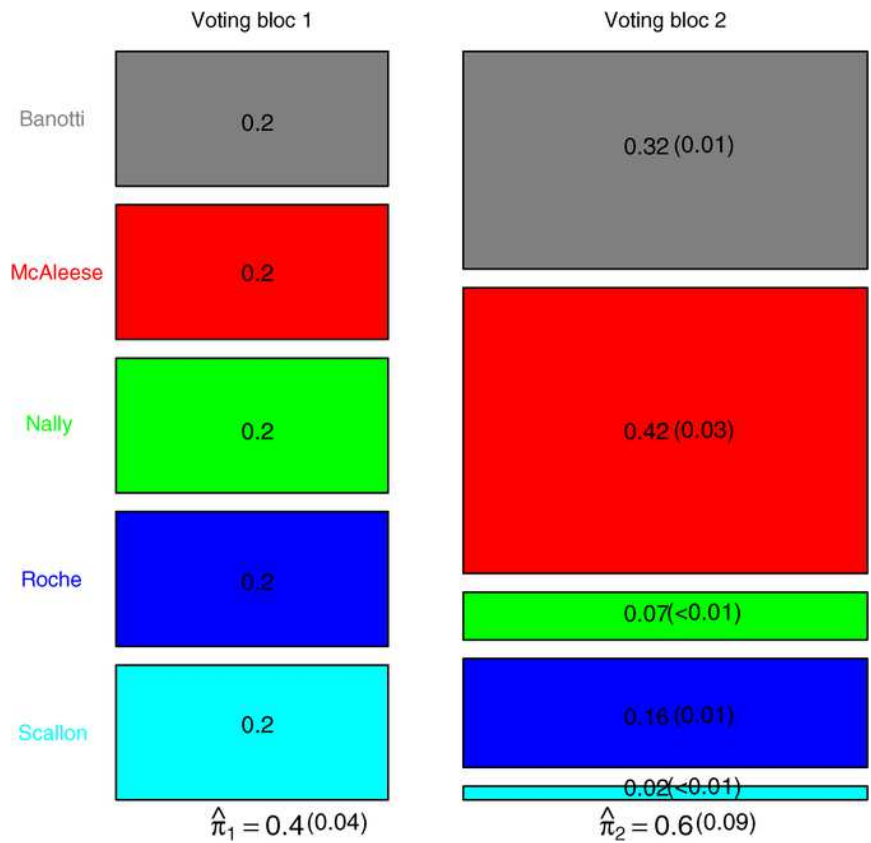

FIG. 4.  *A graphical representation of the maximum likelihood estimates of the support parameters in the two component Plackett–Luce mixture model with a noise component. Each column of the mosaic represents a mixture component or voting bloc—the segments within the columns represent the magnitudes of the support parameters for the candidates within the voting bloc. The width of each column represents the mixture component probability $\pi_k$ of belonging to each voting bloc. Standard errors of the estimates are given in parentheses.*

6.2. *Methods.*   The methods presented in this work are closely related to other statistical methods proposed for the analysis of social science data. Of particular relevance are the following methods.

A mixture model was used in Hill (2001) for the analysis of opinion-changing behavior so that different types of opinion behaviors could be accommodated.

An alternative extension to the mixture model is the mixed membership model for multivariate binary data [Pritchard, Stephens and Peter (2000), Eroshева, Fienberg and Joutard (2007)] and network data [Airoldi et al. (2008)]; such models could be used to study voting blocs in elections that use an approval voting system or votes within a parliamentary system. A mixed membership model for rank data has been developed in Gormley (2006) and Gormley and Murphy (2008) to accommodate mixed membership of voting blocs in STV voting data. The mixed membership model finds more extreme voting blocs than the mixture model because voters are allowed to have partial membership of more than one bloc.

The existence and characterization of voting blocs in the US Senate has been considered in Jakulin and Buntine (2004). Their analysis using exploratory and model-based techniques reveals three Republican and two Democrat voting blocs in the US Senate. An analysis of Asian voting behavior in Tam (1995) showed that Asian voters should not be treated as a monolithic voting bloc. Multidimensional scaling methods were utilized in Holloway (1990) to study voting blocs and their development over time in the United Nations General Assembly.

**7. Discussion and further work.**  This article develops a mixture of experts model for rank data coupled with an efficient hybrid EMM algorithm for model fitting. The model is employed as a model-based clustering technique in which covariate information contributes to the clustering solution. The covariate information contributes to the clustering solution by modeling the component membership parameters of an observation as a generalized linear function of observation covariates. Within the context of rank data, each component in the population is characterized by an appropriate rank data model, the Benter model for rank data.

The MoE model for rank data has been used to establish the presence of four voting blocs in the electorate one month prior to the 1997 Irish Presidential Election. The voting blocs show that the electorate is divided on an ideological and partisan basis with each of the prominant candidates having a bloc of support. We found that age and government satisfaction levels were important social factors in determining membership of the voting blocs.

The dampening parameter in the Benter model indicates that voters were selecting all of their preferences with great care. In fact, the fitted Benter model is very close to the Plackett–Luce model because the dampening parameter is almost equal to one for each choice level. This phenomenon can be explained by the fact that the election only had five candidates. It has previously been shown in Gormley and Murphy (2008a) that in elections with a greater number of candidates the voters select lower preference candidates with less certainty.

The MoE model was able to find a more detailed voting bloc structure than a mixture model analysis of the data. This was because the covariates and the voting behavior both contribute to the structure of the voting blocs, whereas in a mixture model only voting behavior does.

Although the MoE model for rank data proved an appropriate model for Irish voting data, there is still much scope for future research. As stated, a single layer MoE model rather than a hierarchical model was assumed to be sufficient in this context due to the small number of candidates in the presidential race. Irish governmental elections typically involve a large number of candidates and in such a context a single layer model is unlikely to be sufficient. Within the methodology developed in this article, there is no natural metric for selecting the complexity and structure (i.e., the topology) of a hierarchical MoE tree. The use of variational Bayesian methods to estimate a hierarchical MoE model [Bishop and Svensén (2003)] allows both the number of experts and the topology of the associated tree to be determined within a statistically sound framework. The extension to the estimation of a hierarchical MoE model for rank data is a future area of research.

If a multi-layer MoE model is appropriate, the underlying intuition is that the choice process of a voter is a nested process. For example, perhaps a conservatively minded voter chooses a set of conservative candidates first and then from that set chooses a particular candidate. Nested choice models [McFadden (1978), Train (2003)] could be used to model such a choice procedure. These models assume that choices are made in a hierarchical manner; the voters begin with coarse categories which are refined during the choice process. Nested choice models could be extended to nested ranking models using a multi-stage ranking model approach.

The scope of MoE models for rank data lies beyond modeling Irish election data. Many other nations employ preference based voting systems [Gormley and Murphy (2008a)] and the model proposed here can be easily adapted to model such electorates. A number of scholarly societies, including the Institute of Mathematical Statistics and the Royal Statistical Society, use STV voting in their elections and the methods proposed here could be applied to the analysis of their elections. The methodology presented may also be utilized to model other preference data. Irish third level college application choices are analyzed in Gormley and Murphy (2006) using a mixture of Plackett–Luce models and establish the existence of homogeneous groups of applicants. The extension of this research to examine the influence of applicant covariates on third level course choices is a topic of social and educational interest. Additionally, the proposed methodology could be applied to the analysis of customer choice data in marketing applications, where customers express preferences for different products.

Table 5

| Code | Socioeconomic definition |
|------|--------------------------|
| AB | upper middle class & middle class |
| C1 | lower middle class |
| C2 | skilled working class |
| DE | other working class & lowest level of subsistence |
| F50+ | large farmers |
| F50– | small farmers |

## APPENDIX A: DATA SOURCES

The 1997 Irish presidential opinion poll data set was collected by Irish Marketing Surveys and is available through the Irish Elections Data Archive http://www.tcd.ie/Political_Science/elections/elections.html, which is maintained by Professor Michael Marsh in the Department of Political Science, Trinity College Dublin, Ireland.

## APPENDIX B: SOCIOECONOMIC GROUP CODES

Definitions of the socioeconomic group codes used in the opinion poll conducted by Irish Marketing Surveys are provided in Table 5. Further details may be obtained from Millward Brown/Irish Marketing Surveys Limited, www.millwardbrown.com.

## APPENDIX C: THE EMM ALGORITHM FOR THE MIXTURE OF EXPERTS MODEL FOR RANK DATA

Supplementary material [Gormley and Murphy (2008c)] provides a program in C code which may be used to implement the EMM algorithm for the mixtures of experts model for rank data.

When fitting a MoE model for rank data, the EMM algorithm consists of the following steps:

0. Let $h = 0$ and choose initial parameter estimates for the Benter model parameters $\mathbf{p}^{(0)}$, $\underline{\alpha}^{(0)}$ and for the gating network parameters $\boldsymbol{\beta}^{(0)}$.

1. **E step:** Compute the estimates

$$\hat{z}_{ik} = \frac{\pi_{ik}^{(h)} \mathbf{P}\{\underline{x}_i | \underline{p}_k^{(h)}, \underline{\alpha}^{(h)}\}}{\sum_{k'=1}^{K} \pi_{ik'}^{(h)} \mathbf{P}\{\underline{x}_i | \underline{p}_{k'}^{(h)}, \underline{\alpha}^{(h)}\}} \qquad \text{for } i = 1, \dots, M \text{ and } k = 1, \dots, K.$$

Note that by (3.2) the gating network coefficients are defined by

$$\pi_{ik} = \frac{\exp(\underline{\beta}_k^T \underline{w}_i)}{\sum_{k'=1}^{K} \exp(\underline{\beta}_{k'}^T \underline{w}_i)}.$$

2. **M step:** Substituting the $\hat{z}_{ik}$ values obtained in the E step into the complete data log likelihood forms the "Q function"

(C.1)
$$Q = \sum_{i=1}^{M} \sum_{k=1}^{K} \hat{z}_{ik} \left[ \underline{\beta}_{k}^{T} \underline{w}_{i} - \log \left\{ \sum_{k'=1}^{K} \exp(\underline{\beta}_{k'}^{T} \underline{w}_{i}) \right\} \right.$$
$$\left. + \sum_{t=1}^{n_i} \left\{ \alpha_t \log p_{kc(i,t)} - \log \sum_{s=t}^{N} p_{kc(i,s)}^{\alpha_t} \right\} \right],$$

which is maximized with respect to the model parameters during the M step. The dependence of the parameters in $Q$ on estimates from the $h$th iteration of the algorithm is implicit; the notation is suppressed here for reasons of clarity. Due to maximization difficulties, steps from the MM algorithm are embedded in the M step to obtain MLEs of the parameters. Details of the MM algorithm steps are detailed in Appendix C.1. The new maximizing values are $\mathbf{p}^{(h+1)}$, $\underline{\alpha}^{(h+1)}$ and $\boldsymbol{\beta}^{(h+1)}$.

3. If converged, then stop. Otherwise, increment h and return to Step 1.

**C.1. The M step.** The gating network parameters $\boldsymbol{\beta}$ and the Benter model parameters $(\mathbf{p}, \underline{\alpha})$ influence the $Q$ function (C.1) through distinct terms. Hence, the M step reduces to separate maximization problems for each parameter set. Moreover, an ECM algorithm is implemented where the M step consists of a series of conditional maximization steps. Here, the conditional maximizations are with respect to $\underline{p}_1, \ldots, \underline{p}_K$, $\alpha_2, \ldots, \alpha_{N-1}$ and $\underline{\beta}_2, \ldots, \underline{\beta}_K$.

The conditional maximizations are difficult in practice and are therefore implemented using the MM algorithm. This algorithm works by first constructing a surrogate function which minorizes the objective $Q$ function and then maximizing the minorizing surrogate function. This process is iterated leading to a sequence of parameters estimates giving increasing values of the objective $Q$ function.

To construct surrogate functions, mathematical properties of the objective function, or of terms within it, are exploited. One such property is the supporting hyperplane property (SHP) of a convex function. If $f(\theta)$ is a convex function with differential $f'(\theta)$, then the SHP states that

(C.2)              $$f(\theta) \geq f(\theta^{(h)}) + f'(\theta^{(h)})(\theta - \theta^{(h)}).$$

The SHP provides a linear minorizing function which is an ideal candidate for a surrogate function in an optimization transfer algorithm.

Sometimes it may be preferable to form a quadratic or higher order surrogate function. For example, if $f(\theta)$ is a concave function bounding it around

$\theta^{(h)}$, using a quadratic gives

(C.3)
$$f(\theta) \geq f(\theta^{(h)}) + [f'(\theta^{(h)})]^T(\theta - \theta^{(h)})$$
$$+ 1/2(\theta - \theta^{(h)})^T \mathbf{B}(\theta - \theta^{(h)}),$$

where $\mathbf{B}$ is a negative definite matrix such that $H(\theta^{(h)}) > \mathbf{B}$ and $H(\theta^{(h)})$ is the Hessian $d^2 f/d(\theta^{(h)})^2$.

Both these tools are used within the EMM algorithm for rank data as detailed below:

1. *Maximization with respect to the Benter support parameters.* When conditionally maximizing with respect to $p_{kj}$, the dampening parameters are treated as fixed constants, $\bar{\underline{\alpha}}$, equal to the estimates from the previous iteration. Within the $Q$ function (C.1), the term $-\log \sum_{s=t}^{N} p_{kc(i,s)}^{\bar{\alpha}_t}$ is problematic in terms of optimization with respect to $p_{kj}$. However, since the $-\log(\theta)$ function is a strictly convex function, a linear minorizing surrogate function may be obtained via the SHP (C.2), that is,

$$-\log \sum_{s=t}^{N} p_{kc(i,s)}^{\bar{\alpha}_t} \geq -\log \sum_{s=t}^{N} \bar{p}_{kc(i,s)}^{\bar{\alpha}_t} + 1 - \frac{\sum_{s=t}^{N} p_{kc(i,s)}^{\bar{\alpha}_t}}{\sum_{s=t}^{N} \bar{p}_{kc(i,s)}^{\bar{\alpha}_t}},$$

where $\bar{p}_{kj}$ is a constant and, in practice, is the estimate of $p_{kj}$ from the previous iteration. Substituting the nonconstant terms into the objective function, it follows that, up to a constant,

$$Q \geq \sum_{i=1}^{M} \sum_{k=1}^{K} \sum_{t=1}^{n_i} \hat{z}_{ik} \left[ \bar{\alpha}_t \log p_{kc(i,t)} - \left( \frac{\sum_{s=t}^{N} p_{kc(i,s)}^{\bar{\alpha}_t}}{\sum_{s=t}^{N} \bar{p}_{kc(i,s)}^{\bar{\alpha}_t}} \right) \right],$$

which still poses maximization problems. However, implementing the SHP (C.2) of the convex function $f(p) = -p^{\bar{\alpha}_t}$,

$$-p^{\bar{\alpha}_t} \geq -\bar{p}^{\bar{\alpha}_t} - \bar{\alpha}_t \bar{p}^{\bar{\alpha}_t - 1}(p - \bar{p})$$

again provides the surrogate function

$$Q \geq \sum_{i=1}^{M} \sum_{k=1}^{K} \sum_{t=1}^{n_i} \hat{z}_{ik} \left[ \bar{\alpha}_t \log p_{kc(i,t)} - \left\{ \sum_{s=t}^{N} \bar{p}_{kc(i,s)}^{\bar{\alpha}_t} \right\}^{-1} \left\{ \sum_{s=t}^{N} \bar{\alpha}_t \bar{p}_{kc(i,s)}^{\bar{\alpha}_t - 1} p_{kc(i,s)} \right\} \right]$$

up to a constant. Iterative maximization of the surrogate function produces a sequence of $p_{kj}$ values which converge to a maximum of $Q$. Straight forward maximization provides

$$\hat{p}_{kj} = \frac{\omega_{kj}}{\sum_{i=1}^{M} \sum_{t=1}^{n_i} \hat{z}_{ik} \{\sum_{s=t}^{N} \bar{p}_{kc(i,s)}^{\bar{\alpha}_t}\}^{-1} \{\sum_{s=t}^{N+1} \bar{\alpha}_t \bar{p}_{kj}^{\bar{\alpha}_t - 1} \delta_{ijs}\}},$$

where

$$\omega_{kj} = \sum_{i=1}^{M} \sum_{t=1}^{n_i} \hat{z}_{ik} \bar{\alpha}_t \mathbf{1}_{\{j=c(i,s)\}},$$

given that $\mathbf{1}_{\{j=c(i,s)\}}$ is the usual indicator function and

$$\delta_{ijs} = \begin{cases} 1, & \text{if } j = c(i,s) \text{ and } 1 \le s \le n_i, \\ 1, & \text{if } j \ne c(i,l) \text{ for } 1 \le l \le n_i \text{ and } s = N+1, \\ 0, & \text{otherwise.} \end{cases}$$

The methodology presented here is similar to that used when a mixture of Benter models is fitted via the EMM algorithm as detailed in Gormley and Murphy (2008a).

2. *Maximization with respect to the Benter dampening parameters.* In this case the support parameters are treated as constant with $\bar{p}_{kj}$ denoting the estimate from the previous iteration. Returning to the original objective function (C.1), the problematic term $-\log \sum_{s=t}^{M} \bar{p}_{kc(i,s)}^{\alpha_t}$ is a convex function of $\alpha_t$, and employing the SHP (C.2) again gives

$$Q \ge \sum_{i=1}^{M} \sum_{k=1}^{K} \sum_{t=1}^{n_i} \hat{z}_{ik} \left[ \alpha_t \log \bar{p}_{kc(i,t)} + \left( \frac{-\sum_{s=t}^{N} \bar{p}_{kc(i,s)}^{\alpha_t}}{\sum_{s=t}^{N} \bar{p}_{kc(i,s)}^{\bar{\alpha}_t}} \right) \right].$$

As before, this surrogate function still poses optimization problems. However, as $f(\alpha) = -\bar{p}^{\alpha}$ is a concave function, by (C.3),

$$-\bar{p}^{\alpha} \ge -\bar{p}^{\bar{\alpha}} - (\log \bar{p}) \bar{p}^{\bar{\alpha}} (\alpha - \bar{\alpha}) - 1/2 (\alpha - \bar{\alpha})^2 (\log \bar{p})^2,$$

since $H(\bar{\alpha}) > \mathbf{B} = -(\log \bar{p})^2$. This provides the surrogate function

$$\begin{aligned} Q \ge \sum_{i=1}^{M} \sum_{k=1}^{K} \sum_{t=1}^{n_i} \hat{z}_{ik} \Big[ &\alpha_t \log \bar{p}_k c(i,t) \\ &+ \left( \sum_{s=t}^{N} \bar{p}_{kc(i,s)}^{\bar{\alpha}_t} \right)^{-1} \Big\{ \sum_{s=t}^{N} (-\log(\bar{p}_{kc(i,s)}) \bar{p}_{kc(i,s)}^{\bar{\alpha}_t} (\alpha_t - \bar{\alpha}_t) \\ &\qquad\qquad - 1/2 (\alpha_t - \bar{\alpha}_t)^2 (\log \bar{p}_{kc(i,s)})^2 \Big\} \Big] \end{aligned}$$

up to a constant which is a quadratic in $\alpha_t$. Iterative maximization leads to a sequence of $\alpha_t$ estimates which converge to a local maximum of $Q$. Similar methodology is implemented in Gormley and Murphy (2008a) and formulae for $\hat{\alpha}_t$ may be found therein.

3. *Maximization with respect to the gating network parameters.* Maximization of (C.1) with respect to the gating network parameters $\beta_{kl}$ for $k = 2, \ldots, K$ and $l = 0, \ldots, L$ is also not straight forward. The MM algorithm for logistic regression is detailed in Hunter and Lange (2004) and similar methodology is implemented here to achieve MLEs of the gating network parameters.

The $Q$ function, up to a constant, as a function of $\beta$ is

$$(C.4) \qquad Q = \sum_{i=1}^{M} \left[ \sum_{k=1}^{K} \hat{z}_{ik} (\underline{\beta}_k^T \underline{w}_i) - \log \left\{ \sum_{k'=1}^{K} \exp(\underline{\beta}_{k'}^T \underline{w}_i) \right\} \right],$$

since, by definition, $\sum_{k=1}^{K} z_{ik} = 1$. As (C.4) is a concave function, by (C.3), the quadratic function of $\underline{\beta}_k$,

$$Q(\underline{\beta}_k^{(h)}) + Q'(\underline{\beta}_k^{(h)})^T (\underline{\beta}_k - \underline{\beta}_k^{(h)}) + 1/2 (\underline{\beta}_k - \underline{\beta}_k^{(h)})^T \mathbf{B} (\underline{\beta}_k - \underline{\beta}_k^{(h)})$$

minorizes $Q(\underline{\beta}_k)$ at the point $\underline{\beta}_k^{(h)}$ where $\mathbf{B} = -1/4 \sum_{i=1}^{M} \underline{w}_i \underline{w}_i^T$ such that $H(\underline{\beta}_k^{(h)}) > \mathbf{B}$.

Maximizing this minorizing surrogate function gives the iterative update formula

$$(C.5) \qquad \underline{\beta}_k^{(h+1)} = \underline{\beta}_k^{(h)} - \mathbf{B}^{-1} Q'(\underline{\beta}_k^{(h)}),$$

which only requires the inversion of $\mathbf{B}$ once during the iterative algorithm. The similarity with the well-known Newton–Raphson update is apparent — the MM algorithm update (C.5) trades the computational inefficiency of the Newton–Raphson update for an increased number of iterations.

By embedding these MM algorithm steps in the M step of the EM algorithm, a sequence of parameter estimates is produced which converges to (local) MLEs of the Benter model parameters $(\mathbf{p}, \underline{\alpha})$ and of the gating network parameters $\boldsymbol{\beta}$.

**Acknowledgments.** We would like to thank Professor Adrian Raftery, the members of the Center for Statistics and the Social Sciences and the members of the Working Group on Model-based Clustering at the University of Washington for numerous suggestions that contributed enormously to this work.

## SUPPLEMENTARY MATERIAL

**Computing code for "A mixture of experts model for rank data with applications in election studies"** [**Gormley and Murphy (2008c)**] (DOI: 10.1214/08-AOAS178SUPP; .zip). This package contains the data and C programs used to produce the results in this manuscript. The code is explained in the file README.txt and is easily modified to fit the model to alternative data.

SCHOOL OF MATHEMATICAL SCIENCES
UNIVERSITY COLLEGE DUBLIN
IRELAND
E-MAIL: claire.gormley@ucd.ie
        brendan.murphy@ucd.ie